\documentclass[aps,twocolumn,nofootinbib,showpacs,superscriptaddress,pra]{revtex4-1}

\usepackage{calrsfs}
\usepackage{hyperref}
\usepackage{ulem}
\usepackage{graphics}
\usepackage{epstopdf}
\usepackage{bm}
\usepackage{bbm}
\usepackage{amssymb}
\usepackage{epstopdf}
\usepackage{graphicx}
\usepackage[caption=false]{subfig}
\usepackage{amsmath}
\usepackage{color}
\usepackage{array}
\DeclareMathAlphabet{\pazocal}{OMS}{zplm}{m}{n}
\newcommand{\Tt}{\pazocal{T}}
\newcommand{\Ss}{\pazocal{S}}

\newcommand{\rev}[1]{\textcolor{black}{ #1}}

\begin{document}

\title{\rev{Multidimensional quantum-enhanced target detection via spectro-temporal correlation measurements}}





\author{Yingwen \surname{Zhang}}
\affiliation{National Research Council of Canada, 100 Sussex Drive, Ottawa ON Canada, K1A0R6}

\author{Duncan \surname{England}}
\email{Duncan.England@nrc-cnrc.gc.ca}
\affiliation{National Research Council of Canada, 100 Sussex Drive, Ottawa ON Canada, K1A0R6}

\author{Andrei \surname{Nomerotski}}
\affiliation{Physics Department, Brookhaven National Laboratory, Upton, NY 11973, USA}

\author{Peter \surname{Svihra}}
\affiliation{Department of Physics, Faculty of Nuclear Science and Physical Engineering, Czech Technical University, Prague 115 19, Czech Republic}
\affiliation{Department of Physics and Astronomy, School of Natural Sciences, University of Manchester, Manchester M13 9PL, United Kingdom}

\author{Steven \surname{Ferrante}}
\affiliation{Physics Department, Brookhaven National Laboratory, Upton, NY 11973, USA}

\author{Paul \surname{Hockett}}
\affiliation{National Research Council of Canada, 100 Sussex Drive, Ottawa ON Canada, K1A0R6}

\author{Benjamin \surname{Sussman}}
\affiliation{National Research Council of Canada, 100 Sussex Drive, Ottawa ON Canada, K1A0R6}
\affiliation{Department of Physics, University of Ottawa, Ottawa, Ontario, K1N 6N5, Canada}


\begin{abstract}
In this work we investigate quantum-enhanced target detection in the presence of large background noise using multidimensional quantum correlations between photon pairs generated through spontaneous parametric down-conversion. Until now similar experiments have only utilized one of the photon pairs' many degrees of freedom such as temporal correlations and photon number correlations. Here, we utilized both temporal and spectral correlations of the photon pairs and achieved over an order of magnitude reduction to the background noise and in turn \rev{significant} reduction to data acquisition time when compared to utilizing only temporal modes. We believe this work represents an important step in realizing a practical, real-time quantum-enhanced target detection system. The demonstrated technique will also be of importance in many other quantum sensing applications and quantum communications.
	
\end{abstract}

\maketitle

\section{Introduction}
The generation, and detection, of correlated photon pairs has been at the heart of fundamental tests of quantum mechanics~\cite{Aspect1981,Shalm2015} and will be the key to unlocking powerful new technologies~\cite{Jennewein2000,Spring2013,Lloyd2008}. Historically, the theoretical framework~\cite{Bell1964} and seminal demonstrations have considered these correlations from a binary standpoint, with information encoded as 1s and 0s. In reality, the landscape of photon correlation is complex and multidimensional with unbounded degrees of freedom such as time, frequency and angular momentum offering richness beyond two-dimensional polarization. However, single photon detectors struggle to ``image" these multidimensional correlations: Single pixel detectors offer 1D timing measurement, while multi pixel cameras can provide a 2D image but lack the necessary temporal resolution. Through scanning methods, many of these slices can be combined to derive a full picture, but efficient \rev{parallel} measurement of multidimensional correlation has been an outstanding challenge. 

In this work we employ a novel 3D camera technology (the Timepix 3 Camera, TPX3Cam) allowing us to directly measure multidimensional non-classical correlations from a photon pair-source. The 3D camera has 256x256 pixels and, with an appended intensifier, can individually measure the precise position and arrival time of up to $10^7$ photons per second~\cite{Ianzano2018,Nomerotski2019}. We combine this camera with a diffraction grating to make a two-photon spectrometer that, in real time, measures the spectro-temporal correlations produced by the pair source. With an effective measurement technique, these correlations can become a powerful resource for quantum sensing applications. We benchmark this potential by demonstrating their application in quantum-enhanced target detection.

\begin{figure*}[htbp]
	\centering \includegraphics[width=0.8\textwidth]{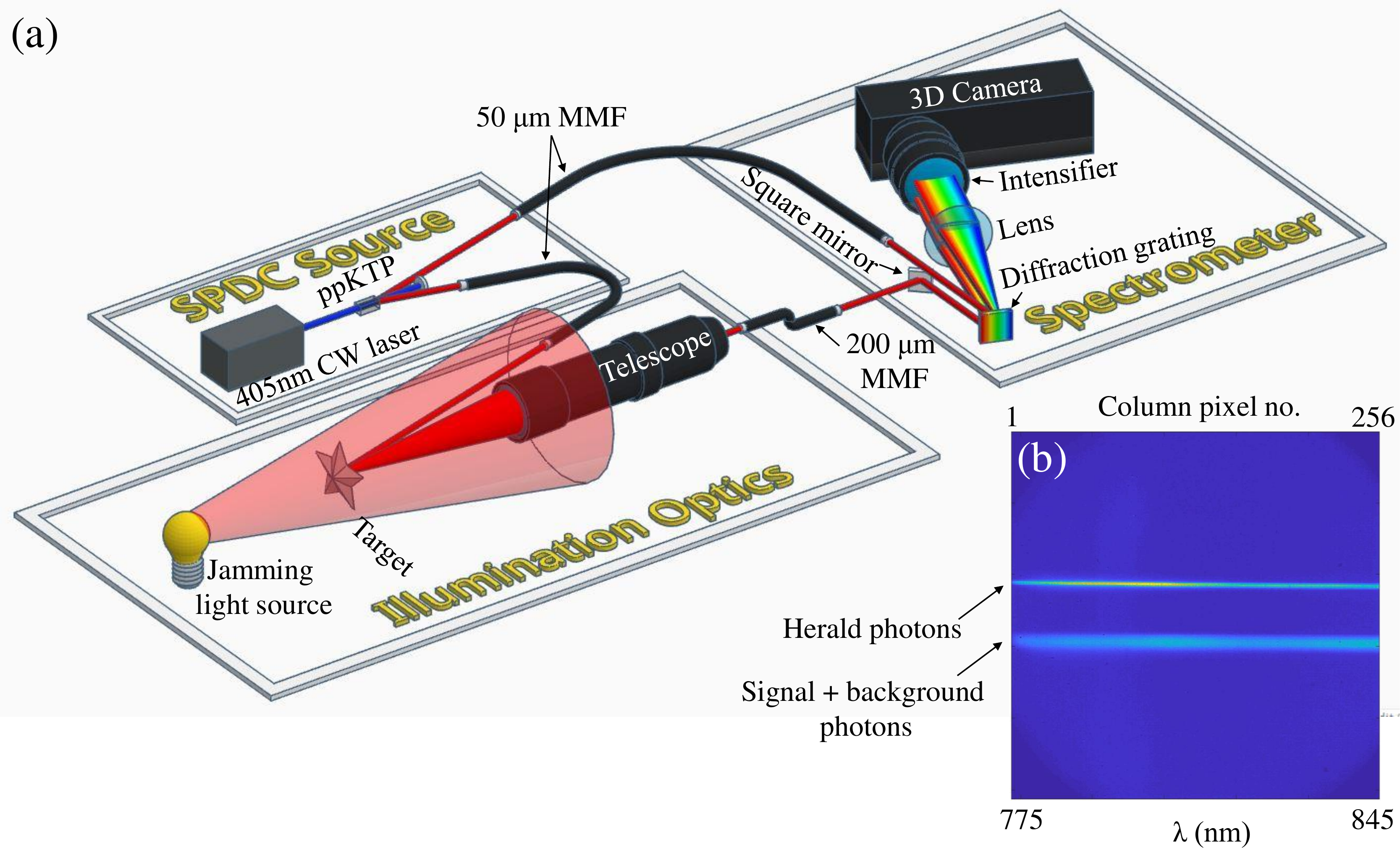}
	\caption{(a) Photon pair Source: Photon pairs correlated in time and wavelength are generated through SPDC by a 405\,nm CW laser pumping a nonlinear crystal. The generated signal and heralding photons are then each coupled into a multi-mode fibre (MMF). Illumination Optics: The signal photon is used to illuminate a target. The diffusely reflected light is collected by a telescope and then coupled into a second fibre. A halogen desk lamp is used to provide broadband background light in the near-infrared to simulate a jamming signal. Spectrometer: The signal and herald photons are vertically displaced when exiting their respective fibres. The photons are then diffracted off a diffraction grating which decomposes the photons into their spectral compositions. The diffracted photons are then focused by a lens onto the 3D camera. (b) Image of the twin beam spectrum captured on the camera.}
	\label{Fig.1}
\end{figure*}

We use the term ``quantum-enhanced target detection" (QTD) as an inclusive term to encompass a range of optical sensing techniques in which quantum light sources are used for remote sensing, offering significant improvements to sensitivity, when compared to using classical light of the same intensity~\cite{Lloyd2008, Tan2008,Shapiro2009,Guha2009,Zhang2013,Aguilar2019,Lopaeva2013,Gregory2019,Ragy2014,England2019,Chang2019,Liu2019}. Typically, these schemes are based on pairs of photons (known as the signal and idler photons) entangled in some degrees of freedom, with the signal photon being sent to illuminate an object while the idler photon is measured locally. The entanglement between signal and idler can be employed to distinguish between signal photons and background photons, providing greater sensitivity in a noisy environment. In the original theoretical proposals, normally referred to as {\it quantum illumination}~\cite{Lloyd2008,Tan2008,Shapiro2009,Guha2009}, the idler photon would need to be stored in a quantum memory until the signal photon returns at which point a joint measurement can be made. Because of the complexity of this scheme, only a limited number of experiments have been performed~\cite{Zhang2013,Aguilar2019}, though always requiring a priori knowledge of the target's position in order to perform the joint measurement.

Due to the complexity in making a joint measurement, many recent experiments demonstrating QTD instead take advantage of the strong correlations between entangled photon pairs. Correlation measurement can be performed without the need for a quantum memory and prior knowledge of the target's distance, and can also effectively distinguish between the signal and background photons even under noisy background conditions. To date, correlation measurements have been performed mainly using just one of the photon pair's degrees of freedom  (e.g. photon number~\cite{Lopaeva2013,Ragy2014,Gregory2019} or time~\cite{England2019,Chang2019,Liu2019}) due to detector limitations. The 3D camera allows us to access both the spectral and the temporal correlations of entangled photon pairs simultaneously. By using two continuous variables rather than one, order-of-magnitude sensitivity improvements are observed. This work is an important step towards realizing QTD in real world applications and the techniques demonstrated in this experiment will also be of great importance to other quantum sensing and quantum communication applications.

\section{Experimental Setup}

\subsection{Photon pair source}
As shown in Fig.~\ref{Fig.1}, a 405\,nm continuous wave (CW) laser is used to pump a periodically-poled potassium titanyl phosphate (ppKTP) crystal to generate pairs of photons known as the signal and idler photon through spontaneous parametric down-conversion (SPDC). \rev{It is well-known that this type of source generates time-frequency entanglement~\cite{MacLean2018}, while our detection apparatus lacks the necessary resolution to prove this, we can nevertheless use the strong correlations in frequency and time to enhance our target detection scheme.}

The signal and idler photons are separated and each coupled into a 50~$\mu$m core diameter multi-mode fibre. To obtain optimal coupling efficiency, imaging lenses (not shown in Fig.~\ref{Fig.1}) are used to image the plane of the crystal onto the fibre, matching the beam diameter of the pump beam on the crystal to the core diameter. The idler photon is used to ``herald" the detection of a signal photon in the 3D camera. The signal photon is used to illuminate a target object (piece of aluminium foil) located at $50$~cm away from the exit of the fiber. Signal photons diffusely reflected by the target are collected by a telescope (45~mm diameter) and then coupled into a multi-mode fibre with 200~$\mu$m core diameter. The collection efficiency of the reflected signal photons is on the order of 1 percent. A halogen desk lamp is used to provide broadband background light in the near infrared to simulate a jamming signal.

\subsection{Coincidence Spectrometer} 
The signal and heralding photons are then sent through the fibres to a customized spectrometer, whose design is similar to that in~\cite{Johnsen2014,Sun2019}, with the two fibers set at a slightly different height with respect to each other. The two beam paths are made parallel by having the signal photons reflect off a square mirror and the heralding photon pass over the mirror. The two beams are then diffracted from grating and the first diffraction order is focused onto the 3D camera using a 2~inch diameter lens. The timing and spectral information of the photons are registered by the 3D camera. When used together with an image intensifier (Cricket from Photonis), the 3D camera is single-photon sensitive and can time-stamp the detection of photons on every pixel. Using this information we are able to deduce the temporal and spectral ($\Tt\&\Ss$) correlations between the signal and heralding photons. A typical image of the twin beam spectrum captured on the camera is seen in Fig.~\ref{Fig.1}(b). The spectral resolution of each beam depends upon the imaging in the spectrometer and size of the fiber core. The herald fiber with a 50\,$\mu m$ core is imaged to a thin stripe resulting in a spectral resolution of 1.3\,nm, while the signal fiber has a larger diameter (200\,$\mu m$) returning a resolution of 3.7\,nm. The spectral range imaged by the camera is approximately $810\pm35$\,nm. A trade-off exists with the smaller core offering higher resolution but larger cores offer higher coupling efficiency. Coupling efficiency is particularly important for the telescope due to the diffuse nature of the reflection for the target, so a larger fibre is chosen in this case.   

\subsection{Timepix3 camera} 
Due to the use of an image intensifier, some care must be taken when analyzing the data from the camera. A single photon is converted into a flash of light by the intensifier, which will often illuminate a cluster of pixels  on the camera. Such a cluster has to be regrouped into a single event through a detection and centroiding algorithm~\cite{Ianzano2018}. Also depending on the signal (flux) detected in each pixel, the measured time of arrival (ToA) of the signal may differ from pixel to pixel even if they are from the same cluster. A larger signal will have a steeper rise time and will therefore cross the discrimination threshold of the camera earlier than smaller signals, giving an earlier ToA. This time-walk effect is also to be corrected in the analysis to improve timing resolution~\cite{Ianzano2018,Zhao2017}. Further details are discussed in the appendix.  

\section{Experimental Results}

\begin{figure}[htbp]
	\centering \includegraphics[width=0.48\textwidth]{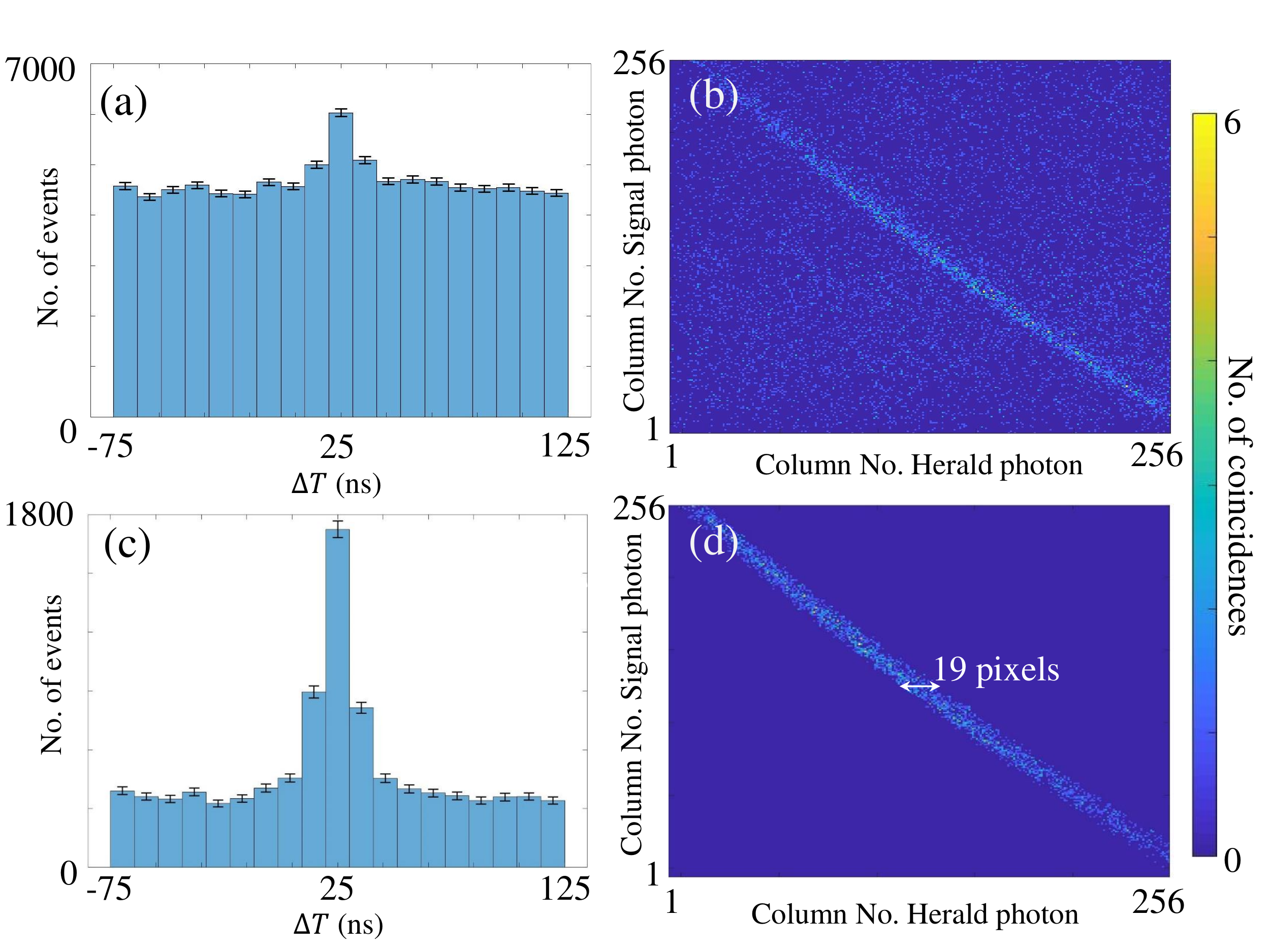}
	\caption{(a) Histogram of 200\,s integration time showing the number of events detected as a function of arrival time difference $\Delta T$ between the signal and heralding photons using only $\Tt$ correlations. The peak is at 25\,ns due to the additional time-of-flight of the signal photon compared to the herald photon. (b) demonstrates $\Tt\&\Ss$ correlation between the herald and signal photons by showing the number of coincidence events between every column (horizontal pixel no.) combination of the herald photon band and the signal photon band from Fig.\ref{Fig.1}(b). (c) shows background reduction in the time-difference histogram when a selection band of 19 pixels wide is placed over the bright $\Ss$ correlation band, as shown in (d), to select out the allowed column (wavelength) combinations.}
	\label{Fig.2}
\end{figure}

For the figures of merit in assessing the performance of the system we use both the signal to background ratio (SBR) and the signal to noise ratio (SNR) defined as:
\begin{align}
	\text{SBR} &= \frac{C_\text{tot}-C_B}{C_B}	\nonumber\\
	\text{SNR} &\rev{= \frac{C_\text{tot}-C_B}{\sigma(C_\text{tot}-C_B)}}
\end{align}
where $C_\text{tot}$ is the total number of measured coincidence events in a given time, $C_B$ is the total background/accidental coincidence events and $\sigma(C_B)$ is the standard deviation in $C_B$. The SNR and SBR become important in different regimes. If the background is known, one can simply subtract the background from the total number of detected photons even if the SBR is very low. In this case, SNR becomes the significant metric and noise, due to shot noise of the background field, it can be reduced simply by a longer data acquisition time (DAT). However, if the background is unknown and fluctuating, then it is not possible to subtract it and SNR no longer improves upon longer DAT. In this case the SBR becomes the significant metric. 

The improvement in SBR and SNR with $\Tt\&\Ss$ correlations is shown in Fig.~\ref{Fig.2}. Here a bright jamming light source is deliberately used to create a large background in order to demonstrate the enhancement in sensitivity, approximately $98\%$ of the detected photons in the signal arm are from the jamming source and environmental background. When treating all the camera pixels in the herald spectrum band as a single detector and those in the signal spectrum band as a second detector, we can determine the SNR and SBR for our QTD system when using only $\Tt$ correlations. Here, the time difference $\Delta T$ between ToA of a signal photon and the ToA of its closest herald photon are determined and a histogram of the number of events vs. $\Delta T$ is shown in Fig.~\ref{Fig.2}(a). The total number of coincidence events are determined by summing up all events that lie within a $\Delta T$ window of  $\tau=20$~ns around the small central peak. The constant base under this small peak are coincidence events between the herald photons and the background and can be determined using (see Appendix)  
\begin{equation}
C_B = \tau\sum_i\sum_jS_iS_j,
\label{m1}
\end{equation}   
where $S_{i(j)}$ is the number of photons detected per second in each column pixel of the signal(herald) photon band, thus $\sum_{i(j)} S_{i(j)}$ is simply the total number of photons detected in the signal(herald) photon band. Multiplying $C_B$ by the data acquisition time (DAT) and subtracting it from the total number of coincidence events will give the number of actual coincidence events between the signal and herald photons. The number of coincidence events between the signal and herald photons in 200~s is determined to be $2130\pm50$ (assuming Poissonian statistics), and the number of background coincidence events is $8610\pm90$. The corresponding SBR and SNR are $0.237\pm0.016$ and \rev{$15.4\pm1.3$ r}espectively. Note that a 20\,ns gating window is used due to the temporal resolution of the camera which is significantly longer than the duration of the photon correlation ($<1$\,ps). Accordingly, the SBR and SNR could be significantly increased if the temporal resolution of the camera were improved.

As the pixel number of every event is also recorded by the camera, one can divide the two spectrum bands in Fig.~\ref{Fig.1}(b) into columns of 1 pixel wide and register the coincidence events of all column combinations between the two bands, we obtain Fig.~\ref{Fig.2}(b). We see that there is a concentration of coincidence events along a central band in Fig.~\ref{Fig.2}(b), an indicator for the presence of $\Ss$ correlations. Converting the column number to wavelength (with the axes in the range of $810\pm35$\,nm), the $\Ss$ correlation band follows the energy conservation relationship:
\begin{equation}
	\lambda_s = \frac{\lambda_p\lambda_h}{\lambda_h-\lambda_p},
\label{Eq2}
\end{equation}   
where the subscripts `s', `h' and `p' stands for the signal, herald and pump respectively and $\lambda_p=405$\,nm. To take advantage of the $\Ss$ correlations, we select out photon detection events only from the column combinations within a selection band following the curve given by equation~(\ref{Eq2}). The width $w$ of the selection band is an important parameter; a small $w$ places a tighter condition on the $\Ss$ correlations thereby reducing accidental background coincidences and giving a better SBR (by up to $\sim 22$ times). However, if $w$ is set too small, then more true coincidences are removed and the SNR suffers. By modeling and experimental optimization (see section~\ref{sec:SE} for details) we see that a width of \rev{$w=19$}\,pixels provides the optimum SNR. The spectral selection is illustrated in Fig.~\ref{Fig.2}(d), and the resulting improvement in the temporal histogram from using only events in this selected region is shown in Fig.~\ref{Fig.2}(c). In this case, the SBR and SNR are \rev{$2.91\pm0.14$ and $34.6\pm1.1$} respectively, giving a SBR enhancement of $12.3pm1.0$ times and SNR enhancement of $2.24\pm0.21$ times compared to when only $\Tt$ correlations were used.

Here we must also mention some of the limitations of QTD seen in this experiment. For the data displayed in Fig.~\ref{Fig.2}, the number of background photons detected in 200~s is $\sim 1.1\times10^7$, in the same time, $\sim 2.5\times10^5$ signal photons are detected. This gives a SBR of 0.023 and SNR of \rev{53} for classical target detection when we use only the signal photons for detection. Although the SBR obtained using $\Tt\&\Ss$ correlations is two orders of magnitude higher, the SNR is \rev{in fact lower than} the classical case. This is due to the low overall detection efficiency ($\sim 1\%$ for the herald photon) and low timing resolution ($\sim 10$\,ns) of the system (see Appendix for details). With better detector technology in the future and a more refined setup, this will no longer be an issue. For an unknown background which cannot simply be subtracted, QTD is still advantageous as SBR becomes the deciding factor.


\subsection{Sensitivity Enhancement} \label{sec:SE}
\begin{figure}[htbp]
	\centering \includegraphics[width=0.45\textwidth]{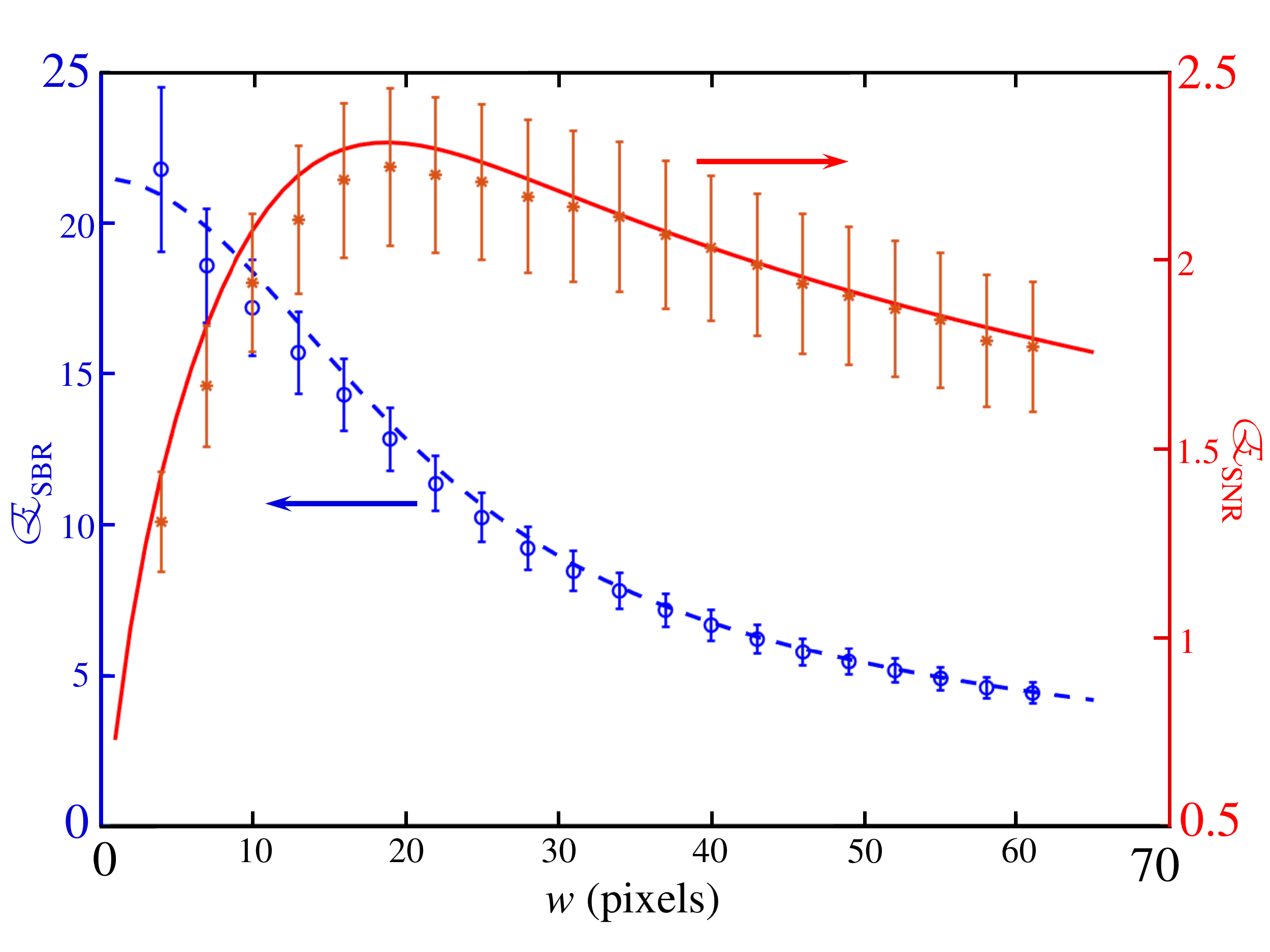}
	\caption{Experimentally measured and theoretically predicted SBR (blue dashed line) and SNR (red solid line) enhancement factors as a function of the width of the $\Ss$ correlation selection band. \rev{Corresponding theoretical curves are from equation~(\ref{Eq6}) with $C=10.6$/s, $S_s=221$/s, $S_h=148$/s $\tau=20$~ns, $N=65536$, $N'=4636$ and $\eta=0.94$. $C$, $S_s$ and $S_h$ are determined from the average of data taken over 200~s. Errorbars are determined using propagation of error from the uncertainties in $C_\text{tot}$ and $C_B$ assuming Poissonian statistics.}}
	\label{Fig.3}
\end{figure}

Previous theoretical~\cite{Lloyd2008} and experimental~\cite{England2019} studies have shown that the sensitivity in QTD is related to the number of available modes. When introducing the spectral degree of freedom, we multiply the number of temporal modes by the number of spectral modes to return the total number of available modes. We would thereby intuitively expect that the SBR would increase by the same factor. The number of measurable spectral modes depends upon the resolution of the two-photon spectrometer and on the width of the selection band. For our spectrometer, with a selection band of \rev{19}\,pixels wide, the signal and herald photons could be split into \rev{13} distinct spectral bins suggesting a potential \rev{13}-fold improvement in SBR compared to $\Tt$ correlations only. This simple estimation is qualitatively close to the experimentally measured improvement of 12.3, showing that the intuitive model is a good approximation. The slight disparity is due to some oversimplifications and a more intricate model, which accurately fits the data, is derived in the appendix. The derived expressions for the SBR and SNR of $\Tt$ correlations ($\text{SBR}_{t}$ and $\text{SNR}_{t}$) and that of $\Tt\&\Ss$ correlations ($\text{SBR}_{ts}$ and $\text{SNR}_{ts}$) are stated below. For the SBR we have:
\begin{align}
\text{SBR}_t &= \frac{C}{N\tau S_sS_h}\nonumber\\
\text{SBR}_{ts} &= \frac{\eta C}{N'\tau S_sS_h},
\label{Eq4}
\end{align}
and for the SNR we have 
\begin{align}
\text{SNR}_t &= \rev{\sqrt{\frac{\text{SBR}_t CT}{\text{SBR}_t+2}}}\nonumber\\
\text{SNR}_{ts} &= \rev{\sqrt{\frac{\text{SBR}_{t}\eta CT}{\text{SBR}_{t}+2N'/(\eta N)}}}.
\label{Eq5}
\end{align}
Here $C$ is the coincidence rate between the signal and herald photons, T is the DAT, $\tau$ the gating time, $S_s$ and $S_h$ are average single photon detection rate per column pixel in the signal and herald spectrum band respectively. $N=256\times256=65536$ is the total number of possible column pixel combinations between the signal and herald spectrum bands. $N'$ is the number of elements in the selection band and is equal to approximately $244w$ for this experiment. Lastly, $\eta$ is an efficiency factor resulting from the reduction of real coincidence events from using the selection band. An expression for $\eta$ can be found in the appendix. An important thing to note is that, unlike SNR, the SBR does not depend on $T$.

From this we get the following expressions for the enhancement factors $\mathcal{E}_{\text{SBR}}$ and $\mathcal{E}_{\text{SNR}}$
\begin{align}
	\mathcal{E}_{\text{SBR}} &= \frac{\text{SBR}_{ts}}{\text{SBR}_t} = \eta\frac{N}{N'} \nonumber\\
	\mathcal{E}_{\text{SNR}} &= \rev{\frac{\text{SNR}_{ts}}{\text{SNR}_t} = \sqrt{\frac{\eta(\text{SBR}_t+2)}{\text{SBR}_{t}+2N'/(\eta N)}}}. 
\label{Eq6}
\end{align}
Equation~(\ref{Eq5}) shows very good agreement with experimental data as seen in Fig.~\ref{Fig.3}. We see that for our setup, $\mathcal{E}_{\text{SBR}}$ reaches a maximum of $\sim 22$ when $w$ approaches zero and $\mathcal{E}_{\text{SNR}}$ reaches a maximum of \rev{$\sim 2.3$ at $w=19$} pixels. \rev{It is interesting to note that as $\text{SBR}_t$ decreases, $\mathcal{E}_{\text{SNR}}$ increases and tends to a maximum value of $\eta\sqrt{N/N'}$ as $\text{SBR}_t \rightarrow 0$. This signifies that the SNR enhancement with multidimensional correlations becomes better with more background noise. Also to note is that when $\text{SBR}_t$ becomes larger than $\frac{2}{\eta}\left|\frac{\eta^2-N'/N}{1-\eta}\right|$, $\mathcal{E}_{\text{SNR}}$ becomes smaller than 1, indicating multidimensional correlations is no longer useful for SNR improvements under low background conditions.}

\begin{figure}[htbp]
	\centering \includegraphics[width=0.45\textwidth]{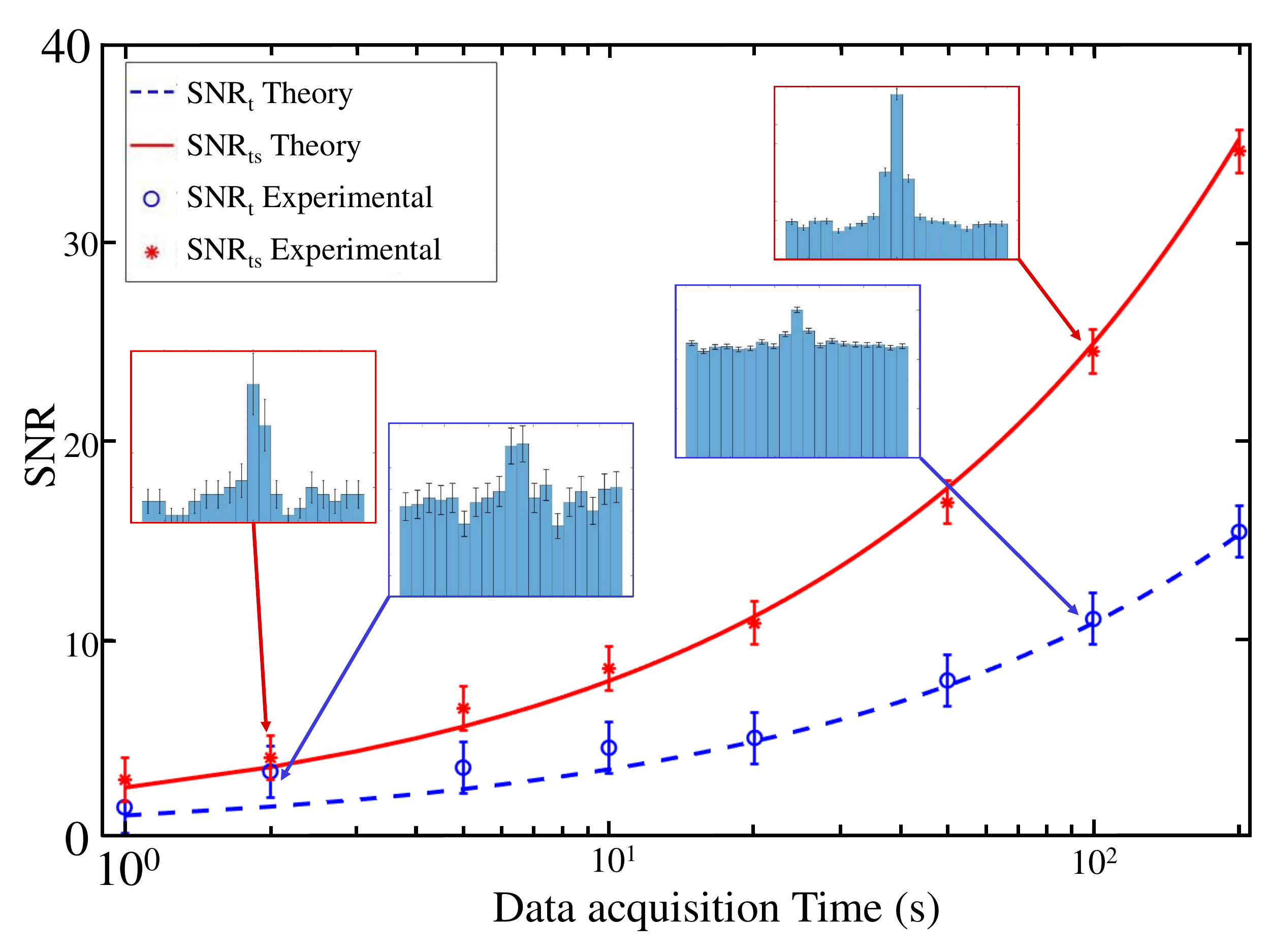}
	\caption{Experimental SNR measured using $\Tt$ correlations (blue circle) and $\Tt\&\Ss$ correlations (red asterisk) as a function of the DAT. For $\Tt\&\Ss$ correlations, the selection band width $w$ is \rev{19} pixels. Time difference histograms for various data points are shown in the insets. Corresponding theoretical curves are from equation~(\ref{Eq5}) with $C=10.6$/s, $S_s=221$/s, $S_h=148$/s $\tau=20$~ns, $N=65536$, $N'=4636$ and $\eta=0.94$. $C$, $S_s$ and $S_h$ are determined from the average of data taken over 200~s. \rev{Errorbars are determined using propagation of error from the uncertainties in $C_\text{tot}$ and $C_B$ assuming Poissonian statistics.}} 
	\label{Fig.4}
\end{figure}

If we set $\text{SNR}_{ts} = \text{SNR}_{t}$, then from equation~(\ref{Eq4}), the DAT can be reduced by a factor given by
\begin{equation}
	\frac{T_t}{T_{ts}} = \rev{\eta^2\left(\frac{C+2N\tau S_sS_h}{\eta C+2N'\tau S_sS_h}\right)} = \mathcal{E}_{\text{SNR}}^2.
\end{equation} 
So for our experiment, a DAT reduction of \rev{5.3} times is possible for \rev{$w=19$}\,pixels when using $\Tt\&\Ss$ correlations. This can be seen in Fig.~\ref{Fig.4}, where we plotted $\text{SNR}_{t}$ and $\text{SNR}_{ts}$ against $T$ for \rev{$w=19$}\,pixels. 

\subsection{Receiver Operating Characteristics} 
For target detection, it is informative to convert photon count rates into a binary ``yes/no'' decision. In this case, we choose a threshold and define a  ``positive" detection as an event where the coincidence counts per unit time exceed this threshold. Due to the intrinsically random nature of photon counting, the number of coincidences fluctuate from shot to shot following a Poissonian distribution~\cite{Mandel1959}; these fluctuations can lead to errors. For example, a target is in place but the coincidence count does not exceed the threshold or a target is not in place but the coincidence count does exceed the threshold resulting in a ``false alarm". The selection of a threshold value therefore becomes important: if the threshold is set too high then we reduce the probability of detection $P_d$, whereas if it is set too low then we increase the probability of a false alarm $P_{fa}$. This leads to an important metric for any receiver --- the so-called receiver operating characteristic (ROC) curve~\cite{Swets1988} --- which compares $P_d$ and $P_{fa}$ as the threshold is varied. The ROC curve is a widely used bench-marking technique in the LiDAR and RADAR community~\cite{Secord2007,Kerekes2008} so is a relevant tool for comparing our technique.

\begin{figure}[htbp]
	\centering \includegraphics[width=0.45\textwidth]{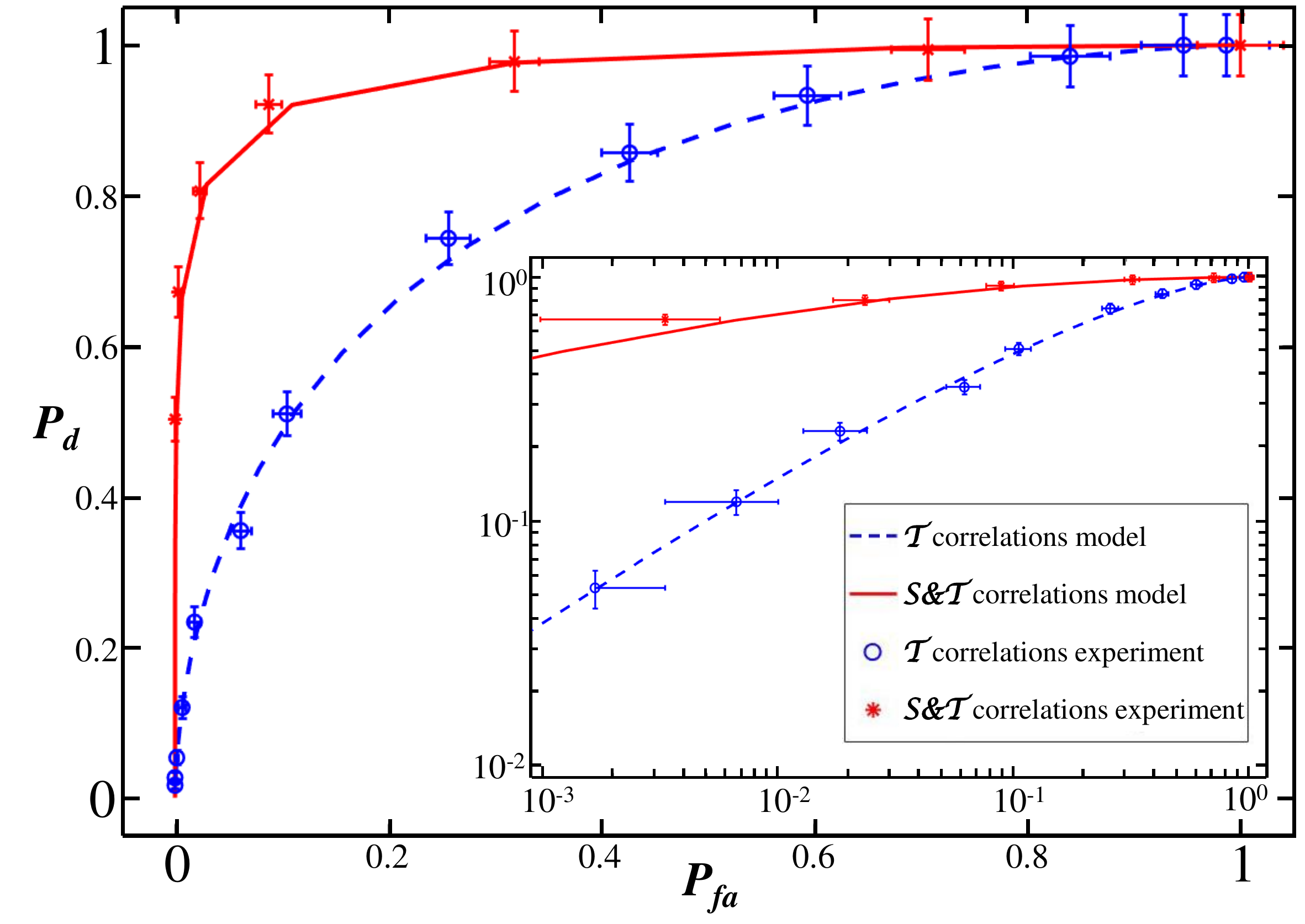}
	\caption{ROC curves for both $\Tt$ (blue circle) and $\Tt\&\Ss$ (red asterisk) correlations with a DAT of 0.5s and $w=14$ pixels. The inset shows the same data in log scale. \rev{Errorbars are determined using propagation of error from the uncertainties in $C_\text{tot}$ and $C_B$ assuming Poissonian statistics.}} 
	\label{Fig.5}
\end{figure}

The ROC curves for $\Tt$ and $\Ss\&\Tt$ correlations are shown in Fig.~\ref{Fig.5}. Here, coincidence data is acquired for 300\,s, and is split into segments of 0.5\,s in duration returning 600 individual data sets. A 20\,ns window centred on the coincidence peak at 25\,ns delay is used to measure the true positives and a second window of the same width centred at 75\,ns is used to measure false alarms from the background. Each of the 600 data sets is analysed for a range of different threshold conditions returning the number of true positives and false alarms for each threshold. The measured data is in good agreement with a simple model based on Poissonian photon statistics and shows a clear improvement when multi-dimensional correlations are used. For example, if we set the threshold such that $P_{fa} = 10^{-3}$, then the detection probability is $\sim 0.5$ for $\Tt\&\Ss$ correlations, but is just $\sim0.04$ for $\Tt$ correlations.  

\section{Discussion}
In this work we have demonstrated QTD using both $\Tt\&\Ss$ correlations in entangled photon pairs generated through SPDC. The experiment demonstrated that it is possible to achieve a 22 times (13dB) improvement to SBR and a \rev{2.3 times (3.6dB)} improvement to SNR when compared to using only $\Tt$ correlations. This could effectively reduce DAT by \rev{$\sim5$} times and significantly improve ROC. The enhancement due to QTD is proportional to the number of measurable modes~\cite{Lloyd2008}. In previous experiments based on only measuring temporal correlations~\cite{England2019}, the number of modes was effectively limited by the temporal resolution of the detectors. By employing both $\Ss$ and $\Tt$ correlations, we have added a second continuous variable to our measurement, and the number of measurable modes is accordingly multiplied allowing us to surpass the sensitivity provided solely by $\Tt$ correlations. In this work, we measured \rev{$13$} distinct spectral modes, further enhancements can be made by improving the spectral and temporal resolution of our two-photon spectrometer. In the future, additional photonic degrees of freedom such as orbital angular momentum correlations~\cite{Mair2001,Leach2010,Dada2011} could be exploited to add further modes. This can be done by sorting on the vertical axis of the 3D camera using an orbital angular momentum sorter~\cite{Berkhout2010,Mirhosseini2013,Larocque2017}, while wavelength is sorted on the horizontal axis.

This work represents an important step towards realizing practical, real-time QTD. The technique used in this experiment for $\Tt\&\Ss$ correlation measurements will also be of use to many other quantum technology applications, such as noise reduction in quantum imaging, and also frequency multiplexing in quantum communication. Presently, due to limitations in detector resolution and efficiency, the SNR improvement compared to classical techniques is modest but, with improved hardware in the near future, the potential advantages offered by multidimensional quantum sensing are vast.

\section*{Appendix}

\subsection*{SBR and SNR derivation} 
We start by first finding an expression for the background coincidence rate when only $\Tt$ correlations are considered. If for the pixel column number $i$ of the upper spectrum band and column $j$ of the lower spectrum band, the number of photons detected per second in each column is $S_i$ and $S_j$ respectively, then statistically for uncorrelated photons, the total background coincidence rate for a gating window of $\tau$ between all possible column combinations would be given by
\begin{equation}
C_B = \tau\sum_i\sum_jS_iS_j.
\label{t1}
\end{equation}   
If we make the assumption that the intensity across the spectrum band is constant then equation~(\ref{t1}) becomes
\begin{equation}
C_B = N\tau S_sS_h,
\label{t2}
\end{equation}  
where $S_s$ and $S_h$ are average photon detection rate per column pixel in the signal and herald spectrum band respectively and $N$ is the total number of possible column combinations, which for the camera is $256\times256 = 65536$. When we apply a selection band to Fig.~2(b) of the main text to include $\Ss$ correlations, we reduce the number of allowed column combinations to $N'$ and the background coincidence rate becomes 
\begin{equation}
C_B'= N'\tau S_sS_h = lw\tau S_sS_h,
\label{t3}
\end{equation}  
where $l=244$ is the length of the selection band and $w$ is the width of the selection band in pixels. 

Assuming Poissonian statistics, the SBR and SNR for $\Tt$ correlations and $\Tt\&\Ss$ correlations are then
\begin{align}
&\text{SBR}_t = \frac{CT}{C_BT} = \frac{C}{N\tau S_sS_h}	\nonumber\\
&\text{SNR}_t = \rev{\frac{CT}{\sqrt{CT+2C_BT}} = \sqrt{\frac{\text{SBR}_t CT}{\text{SBR}_t+2}}}\nonumber\\
&\text{SBR}_{ts} = \frac{C'T}{C'_BT} = \frac{\eta C}{N'\tau S_sS_h}	\nonumber\\
&\text{SNR}_{ts} = \rev{\frac{C'T}{\sqrt{C'T+2C'_BT}} = \sqrt{\frac{\text{SBR}_{t}\eta CT}{\text{SBR}_{t}+2N'/(\eta N)}}}.
\label{ES1}
\end{align}
where $C$ is the coincidence rate between the signal and herald photons, $T$ the data acquisition time (DAT), $\tau$ the gating time and $\eta$ is a detection efficiency factor resulting from throwing away some coincidence events from all correlated photons due to applying a selection band that is narrower than the width of the $\Ss$ correlation band.

If we assume the cross-sectional profile of the $\Ss$ correlation band is a Gaussian of the form $\exp\left[-2(x/\alpha)^2\right]$, with $2\alpha$ being the full width of the Gaussian (at $2\sigma$), then the total number of coincidence events $C'$ inside the selection band is
\begin{align}
C' &\propto \int_{-w/2}^{w/2}\exp{[-2(x/\alpha)^2]} dx \nonumber\\
&\propto \alpha\sqrt{\frac{\pi}{2}} \text{Erf}\left(\frac{w}{\sqrt{2}\alpha}\right)
\end{align}
with $\text{Erf}(x)$ being the error function. Then the efficiency factor $\eta$ is simply
\begin{align}
\eta &= C'/\int_{-\infty}^{\infty}\exp{[-2(x/\alpha)^2]} dx\nonumber\\
&=	\text{Erf}\left(\frac{w}{\sqrt{2}\alpha}\right).
\end{align} 

\begin{figure}[htbp]
	\centering \includegraphics[width=0.45\textwidth]{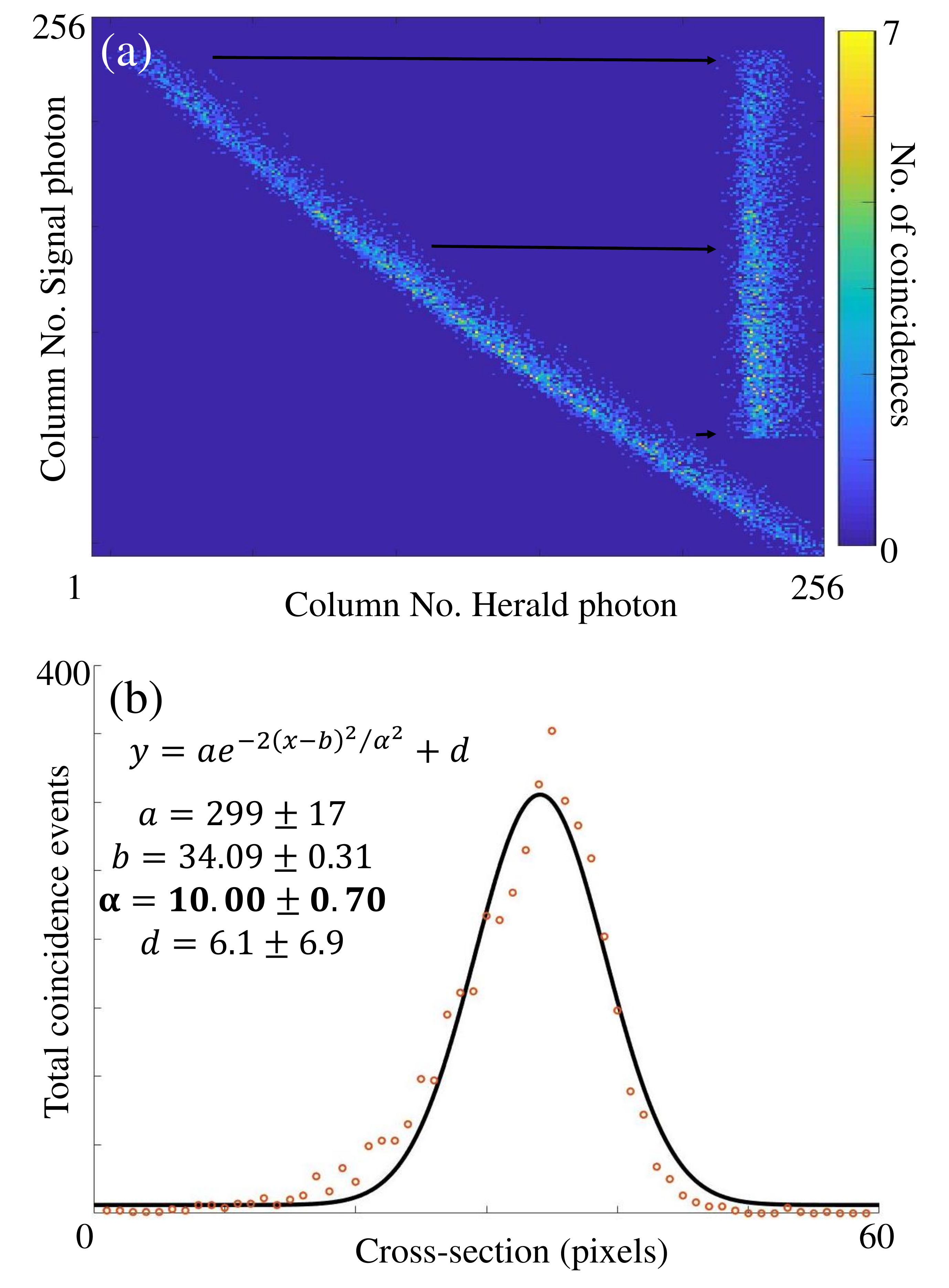}
	\caption{(a) Every row of the spectral correlation band is horizontally shifted such that they form a vertical band. The rows are then added to generate a cross-sectional profile of the correlation band. (b) A Gaussian function is fitted to the cross-sectional profile of the correlation band. The spectral correlation data is taken over 10\,s by directly feeding the signal photons into the 200\,$\mu$m fiber.}
	\label{Fig.s2}
\end{figure}

Through fitting a Gaussian to the cross-section of the spectral correlation band as shown in Fig.~\ref{Fig.s2}, we obtain $\alpha=10.0 \pm 0.7$ pixels.

From equation~(\ref{ES1}), the enhancement factors $\mathcal{E}_{\text{SBR}}$ and $\mathcal{E}_{\text{SNR}}$ are then simply
\begin{align}
	\mathcal{E}_{\text{SBR}} &= \frac{\text{SBR}_{ts}}{\text{SBR}_t} = \eta\frac{N}{N'} \nonumber\\
	\mathcal{E}_{\text{SNR}} &= \frac{\text{SNR}_{ts}}{\text{SNR}_t} = \rev{\sqrt{\frac{\eta(\text{SBR}_t+2)}{\text{SBR}_{t}+2N'/(\eta N)}}}. 
\label{enh}
\end{align}

\subsection*{SBR and SNR of classical vs quantum-enhanced target detection} 
If the photon pair generation rate from our source is $P$, the detection efficiency in the signal and herald arm are $\mu_s$ and $\mu_h$ respectively, the gating time is $\tau$, the number of background photons detected per second in the signal arm is $B$ and assuming the background photons detected in the herald arm is negligible, then with a DAT of $T$, the SBR and SNR for QTD when only temporal correlation is considered is
\begin{align}
\text{SBR}_\text{Q} &= \frac{\mu_s\mu_hPT}{(\mu_s\mu_hP^2+\mu_hPB)\tau T} = \frac{\mu_s}{(\mu_sP+B)\tau}\nonumber\\
\text{SNR}_\text{Q} &= \rev{\frac{\mu_s\mu_hPT}{\sqrt{\mu_s\mu_hPT+2(\mu_s\mu_hP^2+\mu_hPB)\tau T}}}\nonumber\\
&\rev{= \mu_s\sqrt{\frac{\mu_hPT}{\mu_s+2(\mu_sP+B)\tau}}},
\label{SNRQ}
\end{align}  
When related to the previous section, the coincidence rate is $C = \mu_s\mu_hP$, the singles rate in the signal and idler arm are $S_s = (\mu_sP + B)/N$ and  $S_h = (\mu_hP)/N$ respectively.

Now in the case when the background $B$ is much greater than the number of signal photons detected, i.e. $B \gg \mu_sP$ then (\ref{SNRQ}) becomes
\begin{align}
\text{SBR}_\text{Q} &\approx \frac{\mu_s}{B\tau}\nonumber\\
\text{SNR}_\text{Q} &\approx \rev{\mu_s\sqrt{\frac{\mu_hPT}{\mu_s+2B\tau}}}.
\end{align}  

For classical target detection where only the signal photon is used for detection and assuming $B \gg \mu_sP$, then the SBR and SNR is
\begin{align}
\text{SBR}_\text{C} &= \frac{\mu_sPT}{BT} = \frac{\mu_sP}{B}\nonumber\\
\text{SNR}_\text{C} &= \rev{\frac{\mu_sPT}{\sqrt{(\mu_sP+2B)T}} \approx \mu_sP\sqrt{\frac{T}{2B}}}.
\end{align}  

We see that for $\text{SBR}_\text{Q}>\text{SBR}_\text{C}$ would require $\frac{1}{P\tau}>1$ and for $\text{SNR}_\text{Q}>\text{SNR}_\text{C}$ requires $\sqrt{\frac{\mu_h}{P(\mu_s/2B+\tau)}}>1$. For this experiment $P\approx4\times10^6$, $\mu_h\approx0.01$, \rev{$\mu_s\approx3\times 10^{-4}$, $B\approx6\times 10^{4}$} and $\tau=20$\,ns, this gives \rev{$\text{SBR}_\text{Q}/\text{SBR}_\text{C} \approx 13$ and $\text{SNR}_\text{Q}/\text{SNR}_\text{C} \approx 0.3 $}.

\rev{When $\Ss\&\Tt$ correlations are considered, an extra enhancement given by equation (\ref{enh}) is gained, which for a selection band width $w=19$ pixels gives $\text{SBR}'_\text{Q}/\text{SBR}_\text{C} \approx 180 $ and $\text{SNR}'_\text{Q}/ \text{SNR}_\text{C} \approx 0.7$.}

\subsection*{ToA correction}

The shaping of the signal in Timepix3 pixels is dependant on collected energy by each individual channel.
The circuitry defines consistent discharging slope, however, the raising edge reaches the discriminator level earlier for larger energies, creating a so-called time-walk effect.
To counter this effect a lookup table can be generated which gives dependency of $\Delta$ToA (offset from ideal ToA) on ToT (energy measurement).

Normally, the correction is performed using a known time reference by registering external time stamp, however, for the type of data used in this paper such signal cannot be provided.
For this purpose, a previously described algorithm \cite{Ianzano2018} has been utilized.

\begin{figure*}[htbp]
	\centering \includegraphics[width=0.9\textwidth]{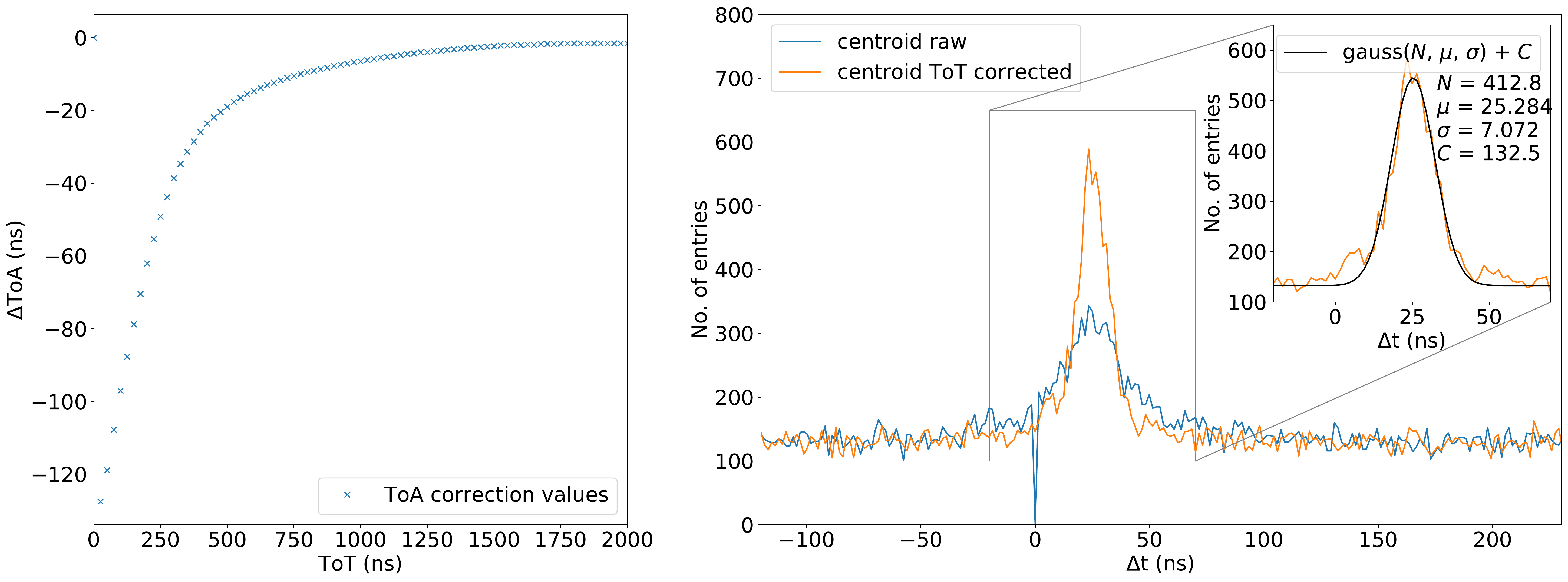}
	\caption{Visualization of applied $\Delta$ToA correction values for different ToT values (left) and its application on a data sample showing $\Delta$t of the entangled photons (right).}
	\label{Fig.s3}
\end{figure*}

As an approximation, the correction uses clustered data to obtain a fixed time reference from a centroid pixel, as all data within the cluster should be due to the same photon, so should have the same arrival time.
Since the measured energy of the centroids varies, a more general dependency on both centroid and pixel ToT is observed and the final correction is obtained using combination of two.
The correction also relies on a fact that for large deposited energies, the time-walk effect asymptotically reaches 0.
Resulting lookup-table values, as well as its application on a data sample is in figure \ref{Fig.s3}.

\subsection*{Specifications of Timepix3 camera}
The TPX3Cam has a resolution of 256x256 pixels with a pixel pitch of 55\,$\mu$m. The single photon sensitivity is provided by the attached intensifier, the cricket from Photonis with Hi-QE Red photocathode, which has a quantum efficiency of approximately $20\%$ between the wavelengths of 600-850\,nm. The timing resolution of the complete system is around 5\,ns (rms) which depends upon many factors including the camera response, the intensifier response, and ToA corrections. More details of the camera can be found in references \cite{Ianzano2018,Nomerotski2019,Zhao2017}. 


\section*{Acknowledgements}
The authors are grateful to Philip Bustard, Rune Lausten, Kate Fenwick, Denis Guay, and Doug Moffatt for technical support and stimulating discussion. We acknowledge support from Defence Research and Development Canada (DRDC).


\bibliography{QIref}

\end{document}